# Effect of Generation of Charged Particles Fluxes by Pulsed Gas Discharge

Georgiy A. Pozdnyakov

*Abstract*— **The paper describes the effect of generation of electron and ion fluxes in a gas discharge, which offers, in particular, to the emergence of "blue jets" and "elves", observed during thunderstorms. An experimental facility modeling these phenomena is described. A possibility of generation of strong-current fluxes of electrons and ions in a straightline gas electric discharge is demonstrated.**

## I. INTRODUCTION

GLOWING of air observed above thunderstorm discharges in the form of "blue jets" and "elves" [1] may be caused, in the author's opinion, by fluxes of electrons and ions generated by thunderstorm discharges. These fluxes are generated owing by two effects accompanying the thunderstorm discharge.

The current channel of the thunderstorm discharge usually has a diameter of about several centimeters, a time of one current pulse with duration of several tens of microseconds, and current amplitude of several $10^4$A. The discharge is accompanied by the release of a large amount of heat in the current channel; therefore, the behavior of air near the current channel can be described by the model of a "strong" or "point" explosion. During the point explosion [2], a centered shock wave is generated, which induces the radial motion of the gas accompanied by a decrease in the gas density at the explosion epicenter. Almost the entire gas is collected near the shock wave front; the gas density distribution is described by the expression $\rho/\rho_2 \propto (r/R)^{3/\gamma - 1}$, where $\rho$ and $\rho_2$ are the gas densities at the point with the coordinate $r < R$ and behind the shock wave front located at the distance R from the explosion epicenter. The maximum rarefaction occurs with a certain delay after the energy release instant. The delay is determined by the duration of the gas spreading phase. During one current pulse of the thunderstorm discharge, the shock wave front (and, correspondingly, the main mass of the gas) can move to a distance of the order of a hundred of millimeters. Thus, intense rarefaction is formed after several tens of microseconds after the discharge beginning, which can lead to generation of fluxes of charged particles if there is a sufficiently strong electric field.

The mechanism of generation of the electron flux in a strong electric field is known as the effect of "runaway of electrons." The loss of energy of the electron due to its collisions with charged particles decreases with increasing energy. A formula for the electron deceleration force $F$ was derived in [3], in the non-relativistic case, this formula is

$$F(q_0) = (4\pi Ne^4 / m_e v^2)\ln(q_0 v/\omega) \qquad (1)$$

Here $q_0$ is the characteristic maximum value of momentum lost by the particle in its collisions, $\omega$ is the mean frequency of motion of electrons in the atom, $N$ is the density of all electrons of the medium, me is the electron mass, e is the electron charge, and $v$ is the electron velocity. Electrons are accelerated if, within the mean free path, they acquire more energy in the external electric field than they lose as a result of deceleration during collisions. Formula (1) ignores the losses due to excitation (the maximum cross section is observed approximately at several electron-volts) and ionization (the maximum cross section is observed approximately at several tens of electron-volts). If electrons acquire the energy of more than several tens of electron-volts within the mean free path, they are accelerated in the external electric field. Obviously, the most effective acceleration is expected to occur on straightline segments of the discharge. These considerations for electrons are also qualitatively valid for ions. Acceleration of the runaway of electrons in a pulsed electric field with sub-nanosecond duration in gases at pressures of the order of the atmospheric value was described in the review [4].

## II. EXPERIMENT

The possibility of generation of charged particles in a straightline gas discharge was experimentally verified in a device schematically shown in Fig. 1. The capacitor C is connected via a controlled discharge device G and an inductor L to the discharge gap between electrodes 1 and 2. The discharge channel 4 passes through the orifices in the dielectric plates 3, as is shown by the arrow in Figure 1. The orifice centers are aligned along a straight line, so that the electric field accelerating the particles is directed along a straight line on the average. The inductance magnitude is chosen from the condition of sufficient duration of the discharge half-period within which necessary rarefaction us

G. A. Pozdnyakov is senior researcher at the Khristianovich Institute of Theoretical and Applied Mechanics SB RAS (e-mail: georg@itam.nsc.ru)

reached on the discharge axis owing to gas spillage induced by discharge initiation. The flux of accelerated charged particles 5 is captured by a collector 6. Between the accelerator (gun) and the collector, there is an earthed grid 7, which cuts off possible discharges between the electrode 1 and the collector; however, this grid is transparent for accelerated particles. The distance between the electrode 2 and the collector 6 is 100 mm, and the collector diameter is 160 mm. The elements 1-7 are located in an evacuated volume. The magnitude of the electric current transferred by accelerated particles is measured by a current transformer 8. The accelerator could operate up to the gas (helium) pressure of 2000 Pa. The greater the atomic weight of the gas (or the mean value for a mixture of gases) in which the discharge occurs, the smaller the maximum pressure at which the collector current was observed.

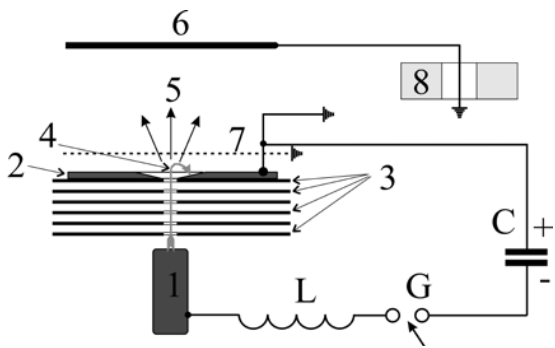

Fig. 1. Scheme of setup. 1 – cathode, 2 – anode, 3 – dielectric plates, 4 – discharge channel, 5 – accelerated particles flux, 6 – collector, 7 – grid, 8 – current transformer.

The glowing of the residue gas induced by accelerated particles is also seen in Fig. 2. Stream of the accelerated particles is strongly expanding. The numbers in the figure are the corresponding accelerator elements.

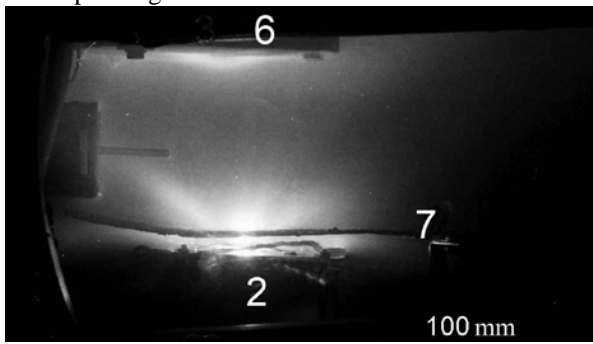

Fig. 2. View of accelerated particles – residue gas interaction. 2 – anode, 6 – collector, 7 – grid.

Figure 3 shows the voltage in the discharge gap 1 and the collector current 2 at a pressure of 0.2 Pa. The results are given for a gum with a distance of 12 mm between the gun electrodes and with an orifice diameter of 1.6 mm. The nominal voltage of capacitor charging is 27 kV, the capacitance is $C$=6.6 μF, and $L$=12 μH.

The collector current has a certain delay with respect to the voltage. Both negative half-waves (electron current) and positive half-waves (ion current) are observed. The ion current ceases after three half-periods.

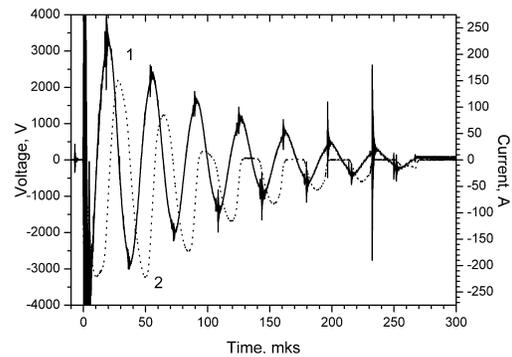

Fig. 3. Voltage (1) and the collector current (2) at a pressure of 0.2 Pa

The dependences of the collector current on time in the case of an evacuated gun (1) and of a gun filled by hydrogen to the pressure $p$=200 Pa (2) are compared in Fig. 4. If the gun is filled by the gas, the ion current becomes drastically smaller than the electron current.

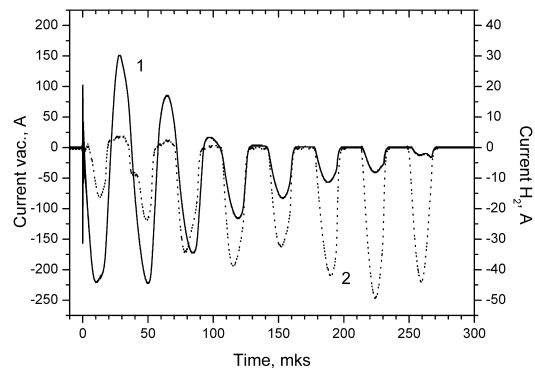

Fig. 4. Collector current at a pressure of 0.2 Pa (1) and 200 Pa (2) ($H_2$)

III. CONCLUSIONS

Generation of fluxes of charged particles can occur in lightnings on straightline segments of the discharge channel, and electrons can acquire the energy up to 2 MeV. Generation of fluxes of high-energy charged particles makes it possible to explain how the electric breakdown of air during thunderstorms occurs at distances much greater than breakdown voltage.

Irradiation of air by powerful fluxes of charged particles generated during thunderstorm discharges induces glowing called "blue jets" and "elves." The efficiency of generation of charged particle fluxes in thunderstorm discharges increases as the pressure decreases (at high altitudes). The effect of ion acceleration is observed at an appreciably lower pressure than the effect of electron acceleration.

Irradiation of air by an electron beam with a high current density leads to the formation of a plasma with a high degree of ionization. It can be assumed that the degree of its imperfection may be fairly significant, especially if air contains water drops, which reduce the plasma temperature. Nonideal plasma can occur when a dense electron beam interaction with the earth surface. Possibly, this is the way how globular lightnings are formed. In 1979, Professor G.E. Norman put forward an idea that the globular lightning is a blob of a strongly nonideal plasma.

Ions accelerated to the energy above 100 keV can enter nuclear reactions when they collide with nuclei of atoms that compose air. Some reports on observations of generation of fluxes of heavy charged particles during thunderstorms are known.

Sources of dense fluxes of electrons similar to the source described above (the polarity of capacitor charging for this case is marked in Fig. 1) possess important distinctive features: workability at comparatively high pressures, large density of the electron current, and simplicity. Significant reduction of the parameters was observed approximately after one hundred operations. Such electron and ion guns can be used, for example, for initiation of chemical reactions, including combustion in various flows [5,6], generation of a dense plasma, and other research and engineering purposes.

In the case of the reverse polarity of the discharge, the device generates a flux of ions. Ion fluxes generated by such a source can be used for research and engineering purposes: for instance, for affecting some surfaces or initiating chemical reactions. Possibly, sources of this type can be used even for initiation and investigation of nuclear reactions, including reactions of synthesis in opposing ion fluxes.

## IV. Acknowledgments

This work was supported by the CRDF (grant # RP0-1393-NO-03) and the Russian Foundation for Basic Research (Grant No. 13-08-00786). The author wishes to express his gratitude to Academician D.D. Ryutov because his lecture inspired the author's idea about the possibility of formation of charged particle beams in thunderstorm discharges and to G.I. Kuznetsov (Budker Institute of Nuclear Physics, SB, RAS) for his kind help in current transformer fabrication. The author gratefully recalls his collaboration with I.A. Golovnov who participated in fabrication and testing of the first model of the accelerator.